\documentclass[twocolumn]{aastex63}
\usepackage{bm}
\usepackage{amsmath}
\usepackage{gensymb}

\newcommand{\gammacool}{\gamma_{\rm cool}}
\newcommand{\gammaacc}{\gamma_{\rm acc}}
\newcommand{\gammacr}{\gamma_{\rm cr}}

\newcommand{\comp}{c/\omega_{\rm p}}
\newcommand{\omp}{\omega_{\rm p}}

\received{xx}
\revised{xx}
\accepted{xx}
\submitjournal{ApJL}

\shorttitle{Polarization Swings during Reconnection-Powered Flares}
\shortauthors{Hosking \& Sironi}
\graphicspath{{./}{figures/}}

\begin{document}

\title{A First-Principle Model for Polarization Swings during Reconnection-Powered Flares}

\author[0000-0002-7958-6993]{David N. Hosking}
\email{david.hosking@physics.ox.ac.uk}
\affiliation{Oxford Astrophysics, Denys Wilkinson Building, Keble Road, Oxford OX1 3RH, UK}
\affiliation{Merton College, Merton Street, Oxford, OX1 4JD, UK}
\author[0000-0002-5951-0756]{Lorenzo Sironi}
\email{lsironi@astro.columbia.edu}
\affiliation{Department of Astronomy and Columbia Astrophysics Laboratory, Columbia University, New York, NY 10027, USA}



\begin{abstract}

We show that magnetic reconnection in a magnetically-dominated fast-cooling plasma can naturally produce bright flares accompanied by rotations in the synchrotron polarization vector. With particle-in-cell simulations of reconnection, we find that flares are powered by efficient particle acceleration at the interface of merging magnetic flux ropes, or ``plasmoids''. The accelerated particles stream through the post-merger plasmoid towards the observer, thus progressively illuminating regions with varying plane-of-sky field direction, and so leading to a rotation in the observed polarization vector. Our results provide evidence for magnetic reconnection as the physical cause of high-energy flares from the relativistic jets of blazars (which recent observations have shown to be frequently associated with polarization rotations), and provide a first-principle physical mechanism for such flares.
\end{abstract}


\section{Introduction} \label{sec:intro}

\begin{figure*}
    \centering
    \includegraphics[scale=0.37]{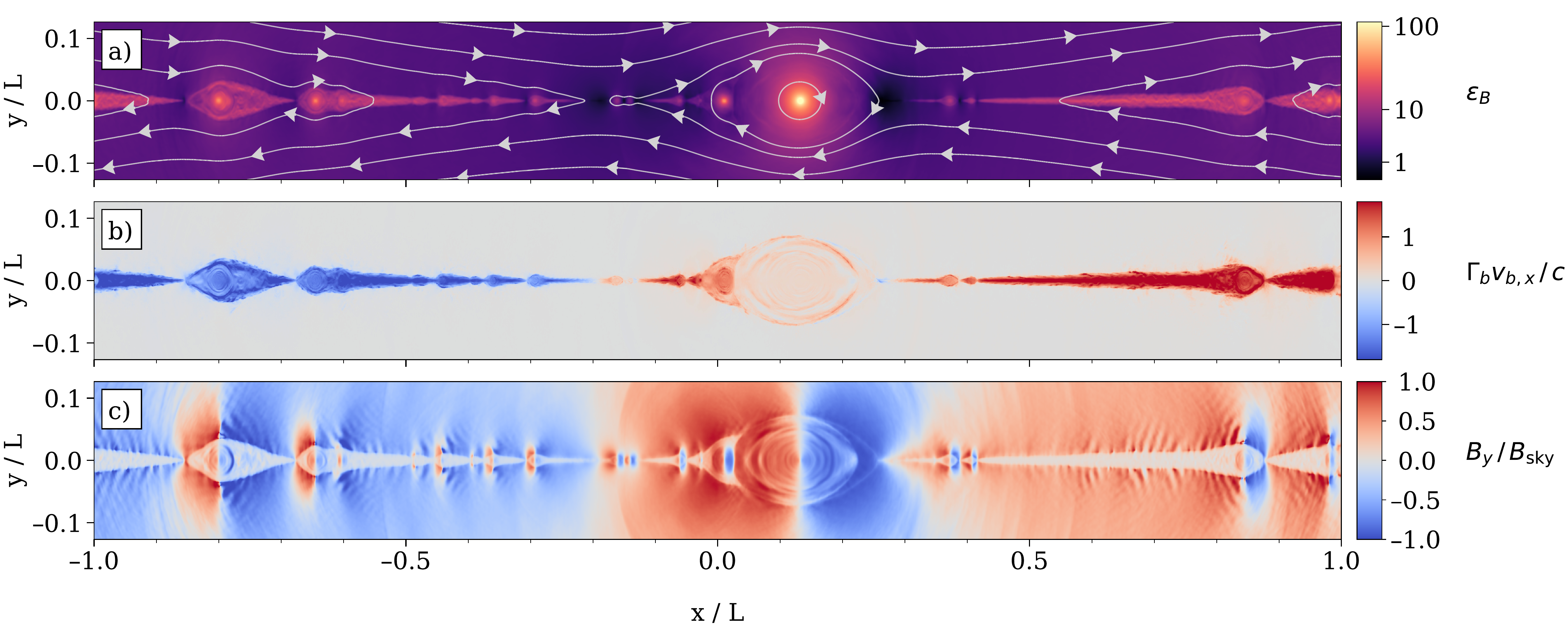}
    \caption{(a) Magnetic energy density normalized by the initial plasma rest-mass energy density, $\epsilon_B = B^2/8\pi n_0 m_e c^2$, with field lines in the $xy$-plane overlaid; magnetic tension drives fast outflows along the reconnection layer. (b) Bulk 4-velocity (in units of $c$), computed by averaging over particle velocities. Since large plasmoids tend to move slower than small ones due to their greater inertia, mergers between a small trailing plasmoid and a large leading plasmoid are common, e.g., at $x \simeq 0$. (c) $B_y/B_{\mathrm{sky}}$, where $B_{\mathrm{sky}}=(B_y^2+B_z^2)^{1/2}$ is the strength of the plane-of-sky field for an observer at $x=\pm\infty$. Since the direction of $\boldsymbol{B}_{\mathrm{sky}}$ rotates within plasmoids, high-energy particles streaming through them naturally induce a PA swing.
}
    \label{fig:fig1}
\end{figure*}

Magnetic reconnection in the relativistic regime \citep{lyutikov_uzdensky_03,lyubarsky_05,comisso_asenjo_14}, where the magnetic energy density is even larger than the particle rest-mass energy density, has been invoked to explain the most dramatic flaring events in astrophysical high-energy sources, most notably the Crab Nebula gamma-ray flares \citep[e.g.][]{cerutti_13a,yuan_16,lyutikov_18} and GeV/TeV  flares of blazars, a class of Active Galactic Nuclei whose relativistic jet points towards Earth \citep[e.g.][]{petropoulou_16,nalewajko_19,mehlhaff_20}.

Our understanding of the physics of relativistic reconnection has recently advanced thanks to fully-kinetic particle-in-cell (PIC) simulations, which have established reconnection as a fast and efficient particle accelerator \citep[e.g.][]{zenitani_01,ss_14,guo_14,werner_16}. As regard to blazars, PIC simulations have demonstrated that reconnection can satisfy all the basic conditions for the emission: efficient dissipation, extended particle distributions, and rough equipartition between particles and magnetic field in the emitting region \citep{sironi_15,petropoulou_19}. In addition, the Doppler-boosted emission of fast ``plasmoids'' (or ``flux ropes'') filled with high-energy particles and magnetic fields --- an essential feature of reconnection \citep{loureiro_07,uzdensky_10} --- can power the ultra-fast bright flares observed at GeV and TeV energies, whose duration can be even shorter than the light-travel time across the black hole that powers the jet \citep{petropoulou_16,christie_19,christie_20}.

Large programs of polarimetric blazar monitoring \citep[e.g. \textit{RoboPol,}][]{angelakis_16}, have recently provided valuable insights into the physics of blazar emission. In some cases, the electric vector position angle (PA) of the polarized emission displays long, smooth and monotonic rotations (or ``swings'') in the optical band, whose amplitudes are as high as hundreds of degrees \citep{marscher_08,marscher_10,abdo_10,larionov_13,aleksic_14a,aleksic_14b,morozova_14,chandra_15}. These are generally associated with multi-wavelength flares and with a temporary decrease in optical polarization degree \citep[hereafter, PD;][]{blinov_15,blinov_16,blinov_18,kiehlmann_16}. 

In this work, we argue that particle acceleration during plasmoid mergers naturally produces bright flares with associated synchrotron PA rotations, providing further evidence for magnetic reconnection as the physical process powering blazar flares. The PA rotations are caused by the apparent rotation of the plane-of-sky magnetic field when the merger-accelerated particles stream towards the observer through the post-merger plasmoid. We demonstrate this mechanism with PIC simulations of relativistic reconnection.
\vspace{0.2in}
\section{Numerical setup}

We employ 2.5D PIC simulations (i.e., 3D vector fields  with translational invariance in $z$) performed with the TRISTAN-MP code \citep{buneman_93, spitkovsky_05}. The in-plane magnetic field is initialized in Harris sheet configuration, with the field along $x$ and reversing at $y=0$ (see Fig. \ref{fig:fig1}). We initiate reconnection by removing the thermal pressure of particles near the center of the sheet at the initial time, as in \citet{sironi_16}. The results we present in Section \ref{sec:results} are obtained at sufficiently late times for the system to have reached a statistically steady state, with no memory of the sheet initialization.

We parameterize the field strength $B_0$ by the magnetization, $\sigma \equiv B_0^2 / 4\pi  n_0 m_e c^2 = \left(\omega_{\rm c} / \omega_{\rm p}\right)^2$, where $\omega_{\rm c} = e B_0 / m_e c$  and $\omega_{\rm p} = \sqrt{4\pi n_0 e^2 / m_e}$ are respectively the Larmor frequency and the plasma frequency for the cold electron-positron plasma outside the layer, with density $n_0$. The Alfv\'{e}n speed is related to the magnetization as $v_A / c = \sqrt{\sigma/\left(\sigma + 1\right)}$; we take $\sigma = 10$ so that $v_A \sim c$, as appropriate for blazar jets. In addition to the reversing in-plane field, we initialize a uniform “guide field” along $-\hat{\boldsymbol{z}}$ with strength $B_g = 0.25 \,B_0$, which helps to provide pressure support to the cores of strongly-cooled plasmoids; we comment on the effect of different guide-field strengths on polarization rotations in Section \ref{sec:discussion}. We resolve the plasma skin depth $c/\omp$ with $5$ cells, and initialize 16 particles in each cell. The numerical speed of light is 0.45 cells/timestep. The box half-length in the $x$-direction is $L\simeq4000\,{\rm cells}= 800\,\comp$. We employ outflow boundary conditions in $x$, while along $y$ two injectors continuously introduce fresh plasma and magnetic flux into the domain \citep[for details see][]{sironi_16,sironi_beloborodov_19}. As opposed to the commonly-adopted double-periodic boundaries, this setup allows us to evolve the system to arbitrarily long times, so we can study the statistical steady state for several Alfv\'enic crossing times (see snapshot at $ct/L\simeq 6.7$ in Fig. \ref{fig:fig1}).

We compute the synchrotron emission \citep[see e.g.][]{cerutti_16} received by an observer at $x=+\infty$ assuming that the radiation is beamed along the particle motion, and including only particles whose velocity falls within a solid angle $\Omega/4\pi=0.03$ around $+\hat{\boldsymbol{x}}$. The corresponding cone is wider than the emission cone of particles with Lorentz factor $\gtrsim \sigma$, which dominate the emission. The inclusion of time retardation would not alter our results other than to reduce the duration of the flare and associated PA swing, and we neglect it for simplicity.

\section{Plasma conditions}\label{sec:pconditions}

Within the blazar class, flat-spectrum radio quasars (FSRQs) generally exhibit the strongest variability and polarized variability \citep{angelakis_16}. For such systems, a hierarchy exists among the timescales (from fast to slow) on which (\textit{i}) particles are accelerated, (\textit{ii}) particles cool and (\textit{iii}) the dynamical time. Equivalently, $\gammacr\gg\gammaacc\gg\gammacool$, where
we define the following characteristic electron Lorentz factors: 
$\gammacr$,  the ``synchrotron burnoff limit'' \citep{dejager_harding_92}, at which synchrotron losses would prohibit further acceleration by the reconnection electric field $E_{\rm rec}=\eta_{\rm rec}B_0$ ($\eta_{\rm rec}\sim 0.1$ \citep[e.g.][]{sironi_16}), i.e.,
\begin{equation}
    e E_{\rm rec}\sim\frac{4}{3}\sigma_{\rm T}\gammacr^2 \,\frac{B_0^2}{8 \pi};\label{gammacrit}
\end{equation} $\gammaacc$, the typical Lorentz factor to which particles are energized, which we estimate by assuming efficient conversion of magnetic energy to particle kinetic energy by reconnection, so $\gammaacc\sim \sigma$;\footnote{This is appropriate for an electron-positron plasma. For an electron-ion plasma, the typical Lorentz factor of electrons energized by reconnection is $\gamma_{\mathrm{acc, }e} \sim \sigma_i m_{i}/m_e\equiv \sigma_e$, where the magnetization $\sigma_i\equiv B_0^2 / 4\pi  n_{0,i} m_i c^2$ is now normalized to the ion rest-mass energy density.} and $\gammacool$, from which particles would cool in a dynamical time $t_{\rm dyn}=2L/c$. Using equation \eqref{gammacrit}, we have
\begin{equation}
\gamma_{\rm cool}\sim\frac{\gamma_{\rm cr}^2}{\alpha_B \eta_{\rm rec} \sqrt{\sigma}} \frac{c/\omega_{\rm p}}{2L},
\end{equation}
where the factor $\alpha_B\sim 3$ reflects the fact that the mean field in plasmoids, where particles spend most of their life, is larger than $B_0$ \citep{sironi_16}.

Typically, in FSRQs $\gammaacc\sim 10^2-10^3\ll \gammacr$ and $\gammacool\sim 0.01-0.1\,\gammaacc\sim 10$ (see Fig.~3 in \citealt{ghisellini_10}; see also \citealt{celotti_08,bottcher_13,sobacchi_20}). In our simulations, we employ $\gammaacc\sim\sigma=10$, $\gammacr=40\gg\gammaacc$ and a system size of $2L\sim 1600\,\comp$, so $\gammacool\sim 1\sim 0.1\,\gamma_{\rm acc}$ as in blazar jets. Therefore, although our runs have smaller $\gammacr$, $\gammaacc$ and $\gammacool$ than blazar jets, they do satisfy the required hierarchy of time and energy scales. With our parameters, particles accelerated by reconnection up to $\gammaacc\sim \sigma$ cool on a timescale $\sim (\gammacool/\gammaacc)\, t_{\rm dyn}\sim 0.2 \,L/c$, which is of the same order as the light crossing time of the largest plasmoids, {with diameter $w\sim 0.2 \,L$} \citep{sironi_16}. 
This is important for our model, since a large PA swing is produced only if the emitting particles do not appreciably cool while moving through the post-merger plasmoid. 

Numerically, we implement synchrotron cooling according to the reduced Landau-Lifshitz model \citep[see][]{vranic_16}. We do not include inverse Compton losses, which may indeed be the most important cooling mechanism in the brightest FSRQs \citep[e.g.][]{celotti_08,bottcher_13}. While the presence of strong cooling is required to allow the sporadic flares associated with plasmoid mergers to dominate the quiescent emission, we do not expect the physical nature of the cooling to impact the occurrence or properties of PA swings, and indeed we have obtained qualitatively similar results when including inverse Compton losses.

\begin{figure}
    \centering
    \includegraphics[scale=0.372]{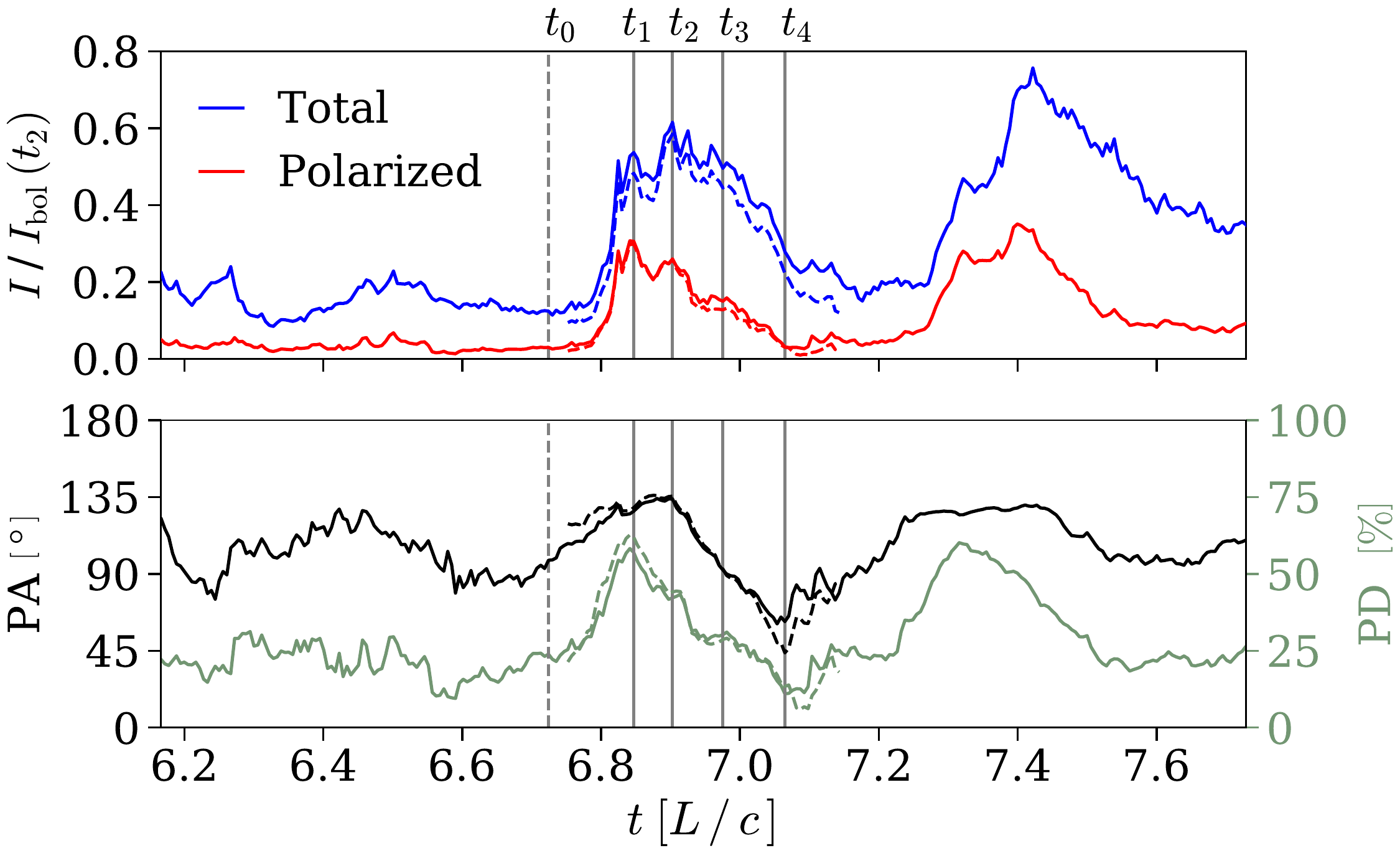}
    \caption{Top: Lightcurves of total and polarized synchrotron intensity in the high-frequency band defined in Fig. \ref{fig:spectrum}, for an observer at $x=+\infty$. Both are normalized to the peak bolometric flux at $t_2$. Bottom: Time series of the PA, measured counterclockwise from $+\hat{\boldsymbol{z}}$, and the PD. Solid lines include the whole simulation domain, dashed lines only the region $0.04\,L\leq x\leq 0.32 \,L$,  which dominates the central flare. }
    \label{fig:lightcurves}
\end{figure}

\begin{figure}
    \centering
    \includegraphics[scale=0.37]{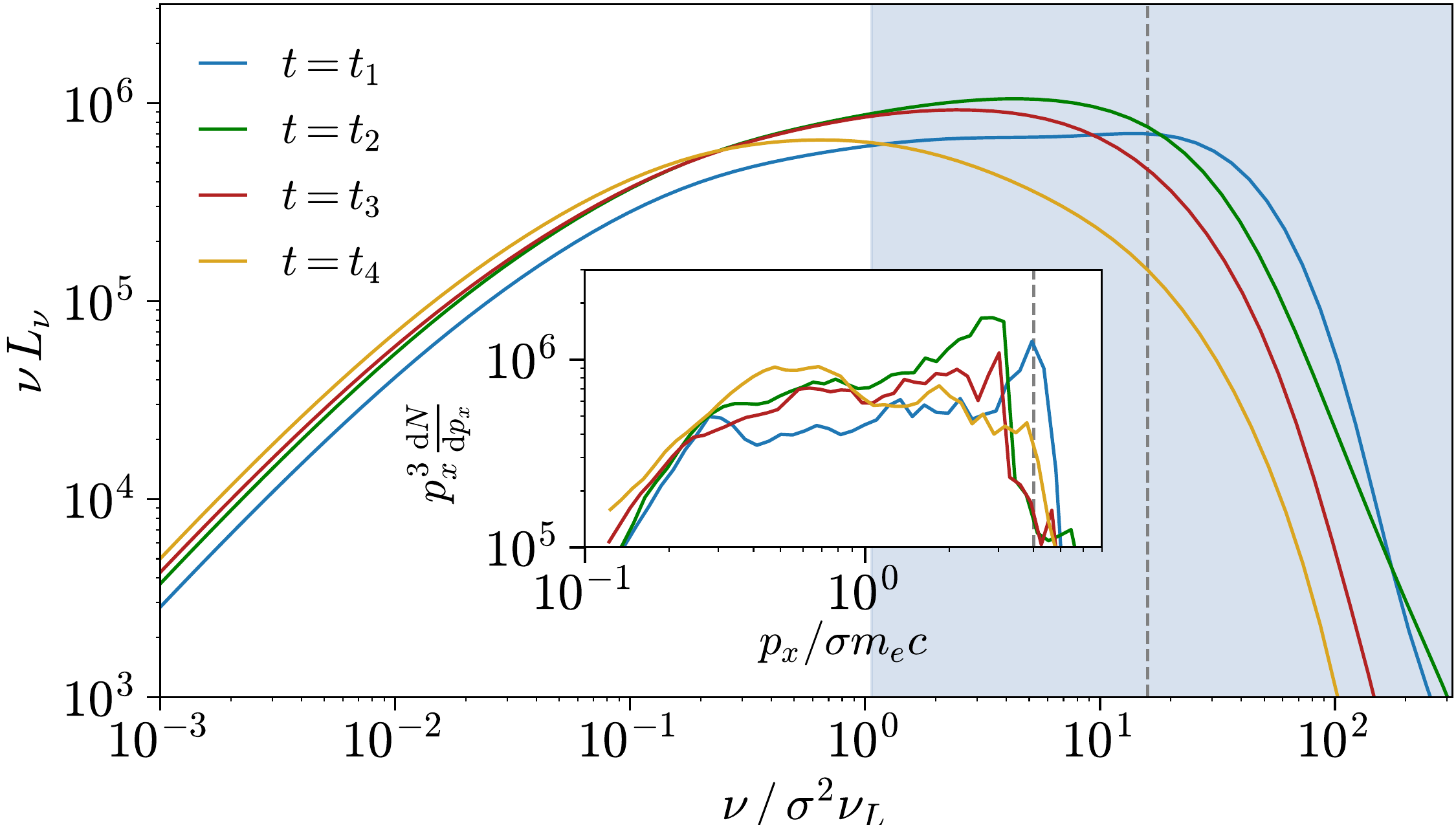}
    \caption{Synchrotron spectrum from the region $0.04\,L\leq x\leq0.32\,L$ at the times $t_i$ defined in the main text (see also Fig. \ref{fig:lightcurves}). The shaded area shows our chosen ``high-frequency band''. 
    The dashed vertical line is at $\nu=\gammacr^2\nu_L$ (the so-called synchrotron burnoff frequency \citep{dejager_harding_92}), where $\nu_{ L}=\omega_{\rm c}/(2\pi)$. 
    Inset: Distribution of $x$-momenta (in units of $\sigma m_e c$) for the particles contributing to the emission. The dashed line is at $p_x = \gammacr m_e c$.}
    \label{fig:spectrum}
\end{figure}

\begin{figure}[ht]
    \centering
    \includegraphics[scale=0.37]{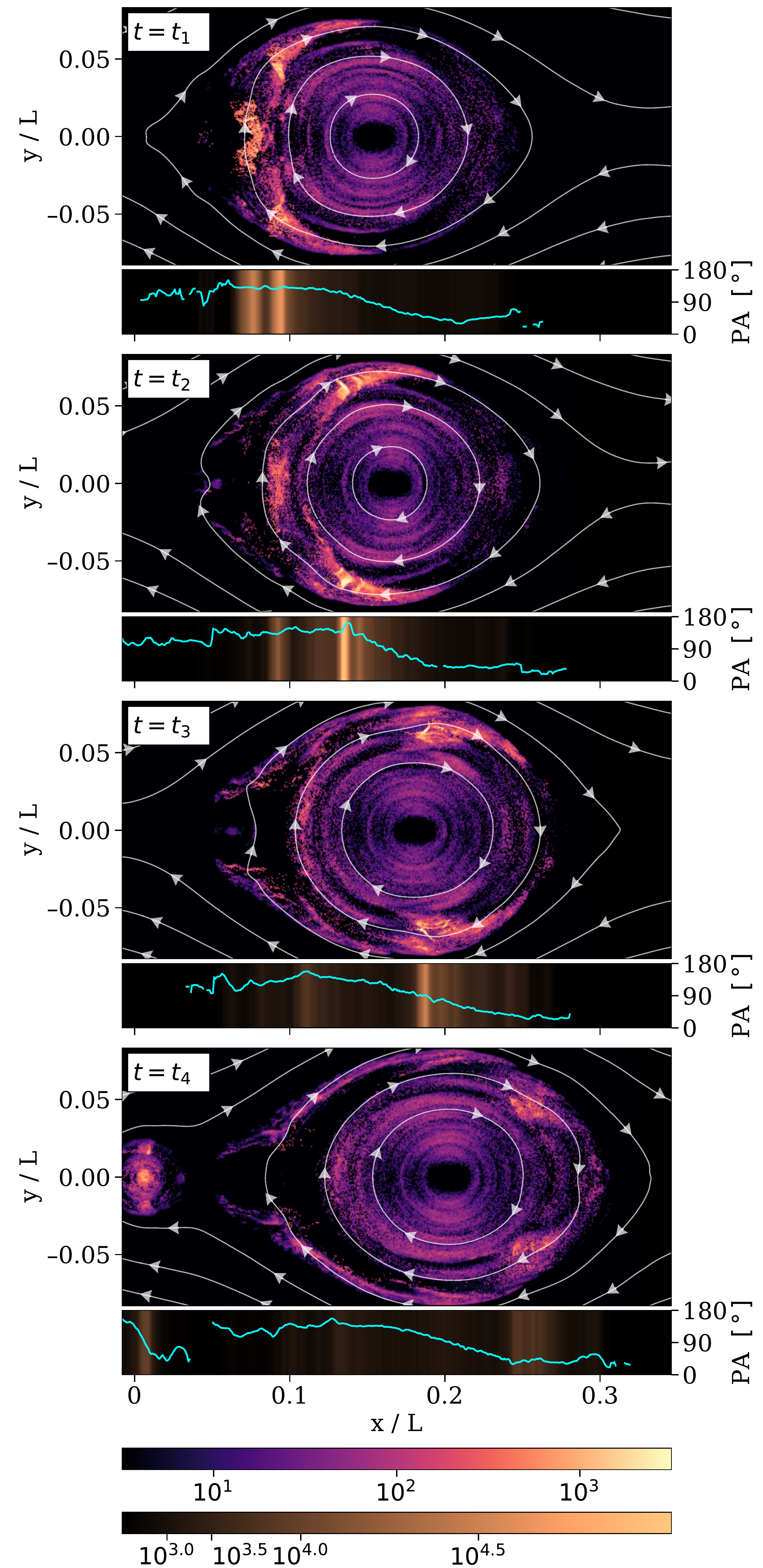}
    \caption{2D bolometric synchrotron emissivity (upper color bar) received by an observer at $x=+\infty$ at the times $t_i$ defined in the main text (see also Fig. \ref{fig:lightcurves}), with field lines in the $xy$-plane overlaid.  Beneath each 2D plot, we show the $y$-integrated high-frequency luminosity (lower color bar) with PA over-plotted in cyan.
As the merger-accelerated particles stream through the plasmoid, they illuminate regions with varying plane-of-sky field, causing the PA swing.
}
    \label{fig:2dmap}
\end{figure}

\section{Results}\label{sec:results}

Figure \ref{fig:lightcurves} shows the  lightcurve of  high-frequency synchrotron emission (defined by the shaded band in Fig. \ref{fig:spectrum}) seen by an observer at $x=+\infty$, together with the PA (measured counterclockwise from $+\hat{\boldsymbol{z}}$) and the PD. The local polarization electric vector is orthogonal to the  plane-of-sky (i.e., $yz$-plane) magnetic field (e.g., $\rm{PA}=90\degree$ if the plane-of-sky field is along the $z$-direction of the guide field). In the prominent flare  from $t \simeq 6.8\, L/c $ to $t \simeq 7.1 \,L/c $ (the first flare in Fig. \ref{fig:lightcurves}), the synchrotron intensity increases by a factor of $\sim 4$, while the PA rotates by $\simeq 70 \degree$. The flare is powered by particles accelerated as a small, fast ``trailing'' plasmoid (located at $-0.02\,L < x < 0.03\,L$ in Fig. \ref{fig:fig1}) merges into a larger and slower ``leading'' plasmoid (located at $0.03\,L < x < 0.23\,L$ in Fig. \ref{fig:fig1}). Most of the flare emission comes from this merger (compare solid and dashed lines in Fig. \ref{fig:lightcurves}). In addition to the pre-merger time $t_0$ displayed in Fig. \ref{fig:fig1}, we define the following times in Fig. \ref{fig:lightcurves}: $t_1$, the peak of  polarized flux and the maximum PD; $t_2$, the peak of total flux and the maximum PA; $t_3$, the midpoint of the polarization swing, when ${\rm PA}=90\degree$; and $t_4$, the minimum PA and PD.

Figure \ref{fig:2dmap} presents the 2D bolometric synchrotron emissivity in the region of the merger, at each time $t_i$. The emission is dominated by merger-accelerated particles that stream around the post-merger plasmoid along its helical magnetic field, whose projection on the $xy$-plane is shown by the grey lines. As the accelerated particles move across the plasmoid, they create an ``emission front'' that sweeps through regions with varying plane-of-sky field direction (see also Fig. \ref{fig:fig1}(c)); in turn, this causes a rotation in the observed PA (see the bright band moving across the lower subpanels of Fig. \ref{fig:2dmap}). 

The flare emission rises on the merger timescale, which is approximately the light-crossing time of the small trailing plasmoid. The subsequent slower decay is due to cooling. A large PA swing requires the cooling time to be on the order of the light-crossing time of the larger, leading plasmoid, so that the emitting particles do not significantly cool before moving across the whole plasmoid. For particles at $\gammaacc\sim \sigma$, the two timescales are indeed comparable, as previously discussed. The size disparity between the two merging plasmoids, and hence the fast rise compared to the slow decay, is essential as the accelerated particles must be localized within a small region in the post-merger plasmoid for their synchrotron emission to be strongly polarized. So, while the two merging plasmoids need to be sufficiently large to energize enough particles to power observable flares, detectable PA rotations only occur if their sizes are somewhat different.

We now describe the temporal evolution of the emission, using Figs.  \ref{fig:spectrum} and \ref{fig:2dmap}. At $t_1$, energetic particles accelerated by the merger are concentrated at the rear of the post-merger plasmoid. At this time (see Fig. \ref{fig:spectrum}), the spectrum has a broad peak, dominated by particles with Lorentz factor $\gamma_{\rm pk}\sim {\rm few}\,\sigma$, and extends up to the synchrotron burnoff frequency $\nu=\gammacr^2 \nu_c$ \citep{dejager_harding_92}, which for our simulation parameters ($\gammacr=40$ and $\sigma=10$) is only marginally greater. For true blazar conditions, where $\gammacr\gg\gammaacc\sim \sigma$, we expect that the emission will extend up to the burnoff limit, but the spectral peak will be dominated by particles with $\gamma_{\rm pk}\sim{\rm few}\,\sigma$, given the steep spectral slopes typically produced by reconnection for blazar conditions \citep{ball_18, petropoulou_19}.

Subsequently, the energetic particles stream along the plasmoid field lines, and cool. By $t_2$, Fig. \ref{fig:spectrum} shows that the peak frequency has dropped by a factor of $\sim 3$ due to cooling --- however, the peak flux is greater, since ongoing  acceleration has increased the number of energetic particles emitting towards the observer.

From $t_2$ to $t_4$, the streaming of energetic particles around the plasmoid induces a PA rotation  of $\simeq 70 \degree$, with the polarization vector rotating from $\simeq 135 \degree$ (measured counterclockwise from $+\hat{\boldsymbol{z}}$), to $\simeq 65 \degree$. Meanwhile, cooling losses reduce both the peak frequency and the peak flux, since particle acceleration has ceased. The polarization degree decreases with time (its maximum is $\simeq 57\%$ at $t_1$) as the polarization signal is diluted by the spatial diffusion and cooling of high energy particles. We note that a drop in optical PD during PA swings has been observed in blazars \citep{blinov_15}. The peak PD and the amplitude of polarization swing are dependent on the chosen frequency band, both being larger when the band is restricted to higher frequencies, which are dominated by the most energetic particles. For example, if we focus only on frequencies higher than the peak frequency at $t_2$,  the maximum PD reaches $\simeq72\%$ \footnote{This is close to the theoretical maximum PD of 75\%, corresponding to monoenergetic particles. Indeed, the inset of Fig.~\ref{fig:spectrum} shows a sharp peak in the momentum spectrum at this time.}, while the amplitude of the PA swing increases to $\simeq90\degree$.

By $t_4$, the PA reaches its minimum, as most of the emitting particles have fully circled the plasmoid. By this time, strong cooling losses have reduced the synchrotron luminosity to pre-merger levels, and merger-accelerated particles no longer dominate the layer-integrated emission (see Fig.  \ref{fig:lightcurves}).

We note that not all flares produced by plasmoid mergers  are accompanied by large PA swings. For example, in the merger of two plasmoids of similar sizes, the emission is not sufficiently localized to produce a strong rotation. Alternatively, when  several small plasmoids merge with the tail of a large one over a timescale similar to the streaming time of accelerated particles, there is no appreciable PA swing because the emitting particles are distributed throughout the post-merger plasmoid (an example is the second flare in Fig. \ref{fig:lightcurves}, at $t\simeq 7.4\, L/c$).

\section{Discussion}\label{sec:discussion}

Polarization swings associated with multi-wavelength flares can help constrain the nature of high-energy emission from blazar jets.
Existing theoretical models of blazar PA swings invoke  geometric effects \citep{marscher_08,marscher_10,lyutikov_17}, stochastic processes in turbulent fields \citep{marscher_14}, or a local magnetic field alteration due to shocks or magnetic instabilities \citep[][see the latter for a reconnection-based model]{zhang_15, nalewajko_17,zhang_18}. Our model belongs to the latter category. We have shown that particles accelerated during mergers of plasmoids produce both a bright flare and a simultaneous PA swing as they stream through the post-merger plasmoid while cooling.

To assess the expected emission frequency, let us assume that the typical Lorentz factor of jet electrons is $\sim 3\times 10^2$, and that the magnetic field strength is $\mathcal{B}=1\,\mathcal{B}_0\,\rm{G}$ \citep{celotti_08,ghisellini_10,bottcher_13}. In reconnection, the mean magnetic energy per particle, $\sigma/2$, is equally divided between accelerated particles and reconnected magnetic fields, giving a mean Lorentz factor of $\sim \sigma /4$ \citep[see][]{sironi_15}. Therefore, we require $\sigma \sim 10^3$ (for an electron-ion plasma, this constraint corresponds to $\sigma_e \sim 10^3$, so the ion-normalized magnetization is $\sigma_i\sim1$). The peak emission is from particles with $\gamma_{\rm pk}\sim 3\, \sigma$ (see Fig.~\ref{fig:spectrum}), whose synchrotron emission frequency is
\begin{equation}
    \nu_{\rm obs}\simeq 5 \times 10^{14} \,\Gamma_{j,1} \left(\frac{\gamma_{\rm pk}}{3\times 10^3}\right)^2 \mathcal{B}_0\,\rm{Hz},
\end{equation}
for a jet with Lorentz factor of $\Gamma_j=10\,\Gamma_{j,1}$, viewed from an angle of  $1/\Gamma_j$ from the axis. So, the peak frequency is indeed expected to fall in the optical band.

We can estimate the timescale for flares according to our model as follows. In the rest frame of the jet, the time for a PA rotation is the time taken for accelerated particles to circle the post-merger plasmoid, which is
\begin{equation}
\label{eq:tjet}
    T_{\rm{jet}}=\frac{w}{v_{\rm front}}\sim\frac{0.2L}{c},
\end{equation}where we have taken $w\sim 0.2\,L$ as the characteristic width of the large post-merger plasmoid, which we assume to move at non-relativistic speeds, and $v_{\rm front}\lesssim c$ is the speed at which the emission front of accelerated particles sweeps through the plasmoid. The resulting timescale matches our results in Fig.~\ref{fig:lightcurves} well (when including time retardation, the PA swing duration will be somewhat shorter). The length $L$ of the layer can be computed by assuming a jet dissipation distance of $\sim 1 \,\rm{pc}$ from the black hole and a jet opening angle of $\sim 0.1 \,\rm{rad}$, so we expect current sheets of size $L=3\times10^{17}L_{17.5} \,\rm{cm}$. The observed timescale will be
\begin{equation}
    T_{\rm{obs}}\sim\frac{1}{\Gamma_j}T_{\rm jet}\simeq 2 \;\Gamma_{j,1}^{-1} L_{17.5} v_{\rm front}^{-1}\;\rm{days}.
\end{equation}
This is around the lower limit of rotation durations detectable by current polarization monitoring programs, owing to limited cadence of observations and the $180\degree$ ambiguity in the PA, but it is consistent as an order-of-magnitude estimate \citep[see, e.g.,][]{blinov_16}. Within our model, we argue that many PA rotations may occur on a timescale that is too short to be detectable, while those that are detected are somewhat rarer events corresponding to particularly large plasmoids, or to the cumulative effect of several consecutive mergers (see below).

In our model, the cooling time of the optical-emitting particles needs to be somewhat shorter than the light crossing time of large plasmoids. In fact, for particles at the peak Lorentz factor $\gamma_{\rm pk}\sim 3\,\sigma$, the cooling time is $\sim 0.06 \, L/c\sim 0.3\,T_{\rm jet}$ (see Section~\ref{sec:pconditions} and Eq.~\ref{eq:tjet}). 
Given this, we  expect that at higher frequencies (e.g., X-rays), the emitting particles will cool even faster, well before circling the post-merger plasmoid.  It is therefore a prediction of our model that there should be no appreciable PA rotations in the X-ray band. At frequencies much below the optical band, the longer particle-cooling time implies that a larger fraction of the layer will simultaneously contribute to the emission, prohibiting significant PA rotations.


Our simulation parameters produce PA swings of $\sim 90\degree$; this is consistent with rotations coincident with bright gamma-ray flares \citep{blinov_18}\footnote{Most observational campaigns only select PA swings $>90\degree$.}. Even larger swings have been observed \citep[e.g.][]{marscher_10,chandra_15}. With regard to this, we point out that in the limit $B_g/B_0\ll1$, our proposed mechanism naturally leads to PA swings of $180\degree$, which is the most commonly observed rotation amplitude \citep{blinov_16}. Indeed, within the reconnection scenario, configurations with weaker guide fields better satisfy the blazar constraints of high efficiency and rough equipartition between particles and fields \citep{sironi_15}. Rotation amplitudes even larger than $180\degree$ can be accounted for in our model as a result of consecutive mergers of a chain of small, trailing plasmoids with a large, leading one. If $B_g/B_0\ll1$, the resulting series of flares and corresponding PA rotations could be cumulatively interpreted as a single flaring episode with continuous PA swing of more than $180\degree$. Indeed, we note that the PA rotation of $\sim 720\degree$ in PKS 1510-089 \citep{marscher_10} occurred over a 50-day period encompassing six gamma-ray flares.


We note that, for a given reconnection geometry and observer's line of sight, our model generally predicts PA swings in one particular direction. While expected to be rarer, swings in the opposite direction \citep[e.g.][]{chandra_15} can be produced in the same geometry by the merger of a \emph{small} leading plasmoid with a \emph{larger} trailing one. Particles accelerated at the merger interface stream backwards through the post-merger plasmoid, so our mechanism occurs in reverse, though with reduced intensity due to the smaller Doppler boosting. More commonly, swings in the opposite direction will be produced by a layer with opposite guide field orientation. Thus, for a given object we do not generally expect a preference for PA rotations in any particular direction, which is consistent with observations.

We conclude with a few remarks and caveats. As compared to the pioneering work by \cite{zhang_18}, who also employed PIC simulations of relativistic reconnection to explain PA rotations in blazars, (\textit{i}) we provide a physically-grounded explanation for PA swings and their association with multi-wavelength flares and reduced optical PD \citep{blinov_15}; (\textit{ii}) we emphasize that PA rotations naturally occur if the observer's line of sight is along the reconnection outflow, which is also required to explain ultra-fast GeV and TeV flares in blazars \citep{giannios_13,petropoulou_16,christie_19,christie_20}; (\textit{iii}) we self-consistently retain the anisotropy of emitting particles.

Our simulations have been initialized with the somewhat idealized Harris sheet reconnection geometry. We do not view this as a major limitation, as the merger discussed in Section \ref{sec:results} occurs $\sim 7$ layer-light-crossing times after the sheet is initialized, so we expect the state of the system at this time to be insensitive to choices at initialization. We acknowledge, however, that the global geometry of reconnection layers in the jet may be more complicated than the planar setup considered here. Understanding the statistics of polarization rotations resulting from different layer properties will require a further detailed study, though we speculate that polarimetric measurements may ultimately be useful to distinguish whether current sheets are introduced at the jet base (in so-called “striped jets”, e.g., \citealt{giannios_uzdensky_19}) or whether they are produced by the nonlinear development of MHD instabilities, like the kink mode \citep[e.g.,][]{bodo_20}.


Although we have presented results for a pair plasma,  we expect that our model will also hold in relativistic electron-proton and electron-positron-proton reconnection, since leptons still pick up a significant fraction of the dissipated magnetic energy \citep{rowan_17,rowan_19,werner_18,petropoulou_19}. We defer an investigation of the more complex 3D case to future work, though we note that mergers of plasmoids are observed in 3D simulations of relativistic reconnection \citep[e.g.][]{ss_14,guo_14,werner_17,sironi_beloborodov_19}, so we expect our model to be applicable to the 3D case.

Finally, we reiterate that we have not included inverse Compton losses, which may indeed be the most important cooling mechanism in the brightest FSRQs. While we do not expect the physical nature of the cooling to impact the occurrence or properties of PA  swings, a self-consistent inclusion of both inverse Compton and synchrotron cooling will be required for direct comparison to the multi-wavelength polarimetric signatures of blazars.

\acknowledgments
We thank D. Blinov, L. Comisso, D. Giannios, E. Sobacchi, F. Tavecchio and H. Zhang for useful feedback. This research was carried out in part during the 2019 Summer School at the Center for Computational Astrophysics, Flatiron Institute. The Flatiron Institute is supported by the Simons Foundation. DNH is supported by a UK STFC studentship. LS acknowledges support from the Sloan Fellowship, the Cottrell Scholar Award, DoE DE-SC0016542, NSF ACI-1657507, NASA ATP NNX17AG21G and NSF PHY-1903412. The simulations have been performed at Columbia (Habanero and Terremoto), and with NERSC (Cori) resources.

\bibliography{blob_ApJL}{}

\begin{thebibliography}{}
\expandafter\ifx\csname natexlab\endcsname\relax\def\natexlab#1{#1}\fi
\providecommand{\url}[1]{\href{#1}{#1}}
\providecommand{\dodoi}[1]{doi:~\href{http://doi.org/#1}{\nolinkurl{#1}}}
\providecommand{\doeprint}[1]{\href{http://ascl.net/#1}{\nolinkurl{http://ascl.net/#1}}}
\providecommand{\doarXiv}[1]{\href{https://arxiv.org/abs/#1}{\nolinkurl{https://arxiv.org/abs/#1}}}

\bibitem[{{Abdo} {et~al.}(2010){Abdo}, {Ackermann}, {Ajello}, {Axelsson},
  {Baldini}, {Ballet}, {Barbiellini}, {Bastieri}, {Baughman}, {Bechtol},
  {et~al.}}]{abdo_10}
{Abdo}, A.~A., {Ackermann}, M., {Ajello}, M., {et~al.} 2010, Natur, 463, 919,
  \dodoi{10.1038/nature08841}

\bibitem[{{Aleksi{\'c}} {et~al.}(2014{\natexlab{a}}){Aleksi{\'c}}, {Ansoldi},
  {Antonelli}, {Antoranz}, {Babic}, {Bangale}, {Barres de Almeida}, {Barrio},
  {Becerra Gonz{\'a}lez}, {Bednarek}, {et~al.}}]{aleksic_14a}
{Aleksi{\'c}}, J., {Ansoldi}, S., {Antonelli}, L.~A., {et~al.}
  2014{\natexlab{a}}, A\&A, 567, A41, \dodoi{10.1051/0004-6361/201323036}

\bibitem[{{Aleksi{\'c}} {et~al.}(2014{\natexlab{b}}){Aleksi{\'c}}, {Ansoldi},
  {Antonelli}, {Antoranz}, {Babic}, {Bangale}, {Barres de Almeida}, {Barrio},
  {Becerra Gonz{\'a}lez}, {Bednarek}, {et~al.}}]{aleksic_14b}
---. 2014{\natexlab{b}}, A\&A, 569, A46, \dodoi{10.1051/0004-6361/201423484}

\bibitem[{{Angelakis} {et~al.}(2016){Angelakis}, {Hovatta}, {Blinov},
  {Pavlidou}, {Kiehlmann}, {Myserlis}, {B{\"o}ttcher}, {Mao}, {Panopoulou},
  {Liodakis}, {et~al.}}]{angelakis_16}
{Angelakis}, E., {Hovatta}, T., {Blinov}, D., {et~al.} 2016, MNRAS, 463, 3365,
  \dodoi{10.1093/mnras/stw2217}

\bibitem[{{Ball} {et~al.}(2018){Ball}, {Sironi}, \& {{\"O}zel}}]{ball_18}
{Ball}, D., {Sironi}, L., \& {{\"O}zel}, F. 2018, ApJ, 862, 80,
  \dodoi{10.3847/1538-4357/aac820}

\bibitem[{{Blinov} {et~al.}(2015){Blinov}, {Pavlidou}, {Papadakis},
  {Kiehlmann}, {Panopoulou}, {Liodakis}, {King}, {Angelakis}, {Balokovi{\'c}},
  {Das}, {et~al.}}]{blinov_15}
{Blinov}, D., {Pavlidou}, V., {Papadakis}, I., {et~al.} 2015, MNRAS, 453, 1669,
  \dodoi{10.1093/mnras/stv1723}

\bibitem[{{Blinov} {et~al.}(2016){Blinov}, {Pavlidou}, {Papadakis}, {Hovatta},
  {Pearson}, {Liodakis}, {Panopoulou}, {Angelakis}, {Balokovi{\'c}}, {Das},
  {et~al.}}]{blinov_16}
{Blinov}, D., {Pavlidou}, V., {Papadakis}, I.~E., {et~al.} 2016, MNRAS, 457,
  2252, \dodoi{10.1093/mnras/stw158}

\bibitem[{{Blinov} {et~al.}(2018){Blinov}, {Pavlidou}, {Papadakis},
  {Kiehlmann}, {Liodakis}, {Panopoulou}, {Angelakis}, {Balokovi{\'c}},
  {Hovatta}, {King}, {et~al.}}]{blinov_18}
{Blinov}, D., {Pavlidou}, V., {Papadakis}, I., {et~al.} 2018, MNRAS, 474, 1296,
  \dodoi{10.1093/mnras/stx2786}

\bibitem[{{Bodo} {et~al.}(2020){Bodo}, {Tavecchio}, \& {Sironi}}]{bodo_20}
{Bodo}, G., {Tavecchio}, F., \& {Sironi}, L. 2020, arXiv e-prints,
  arXiv:2006.14976

\bibitem[{{B{\"o}ttcher} {et~al.}(2013){B{\"o}ttcher}, {Reimer}, {Sweeney}, \&
  {Prakash}}]{bottcher_13}
{B{\"o}ttcher}, M., {Reimer}, A., {Sweeney}, K., \& {Prakash}, A. 2013, ApJ,
  768, 54, \dodoi{10.1088/0004-637X/768/1/54}

\bibitem[{{Buneman}(1993)}]{buneman_93}
{Buneman}, O. 1993

\bibitem[{{Celotti} \& {Ghisellini}(2008)}]{celotti_08}
{Celotti}, A., \& {Ghisellini}, G. 2008, MNRAS, 385, 283,
  \dodoi{10.1111/j.1365-2966.2007.12758.x}

\bibitem[{{Cerutti} {et~al.}(2016){Cerutti}, {Mortier}, \&
  {Philippov}}]{cerutti_16}
{Cerutti}, B., {Mortier}, J., \& {Philippov}, A.~A. 2016, MNRAS, 463, L89,
  \dodoi{10.1093/mnrasl/slw162}

\bibitem[{{Cerutti} {et~al.}(2013){Cerutti}, {Werner}, {Uzdensky}, \&
  {Begelman}}]{cerutti_13a}
{Cerutti}, B., {Werner}, G.~R., {Uzdensky}, D.~A., \& {Begelman}, M.~C. 2013,
  ApJ, 770, 147, \dodoi{10.1088/0004-637X/770/2/147}

\bibitem[{{Chandra} {et~al.}(2015){Chandra}, {Zhang}, {Kushwaha}, {Singh},
  {Bottcher}, {Kaur}, \& {Baliyan}}]{chandra_15}
{Chandra}, S., {Zhang}, H., {Kushwaha}, P., {et~al.} 2015, ApJ, 809, 130,
  \dodoi{10.1088/0004-637X/809/2/130}

\bibitem[{{Christie} {et~al.}(2019){Christie}, {Petropoulou}, {Sironi}, \&
  {Giannios}}]{christie_19}
{Christie}, I.~M., {Petropoulou}, M., {Sironi}, L., \& {Giannios}, D. 2019,
  MNRAS, 482, 65, \dodoi{10.1093/mnras/sty2636}

\bibitem[{{Christie} {et~al.}(2020){Christie}, {Petropoulou}, {Sironi}, \&
  {Giannios}}]{christie_20}
---. 2020, MNRAS, 492, 549, \dodoi{10.1093/mnras/stz3265}

\bibitem[{{Comisso} \& {Asenjo}(2014)}]{comisso_asenjo_14}
{Comisso}, L., \& {Asenjo}, F.~A. 2014, \prl, 113, 045001,
  \dodoi{10.1103/PhysRevLett.113.045001}

\bibitem[{{de Jager} \& {Harding}(1992)}]{dejager_harding_92}
{de Jager}, O.~C., \& {Harding}, A.~K. 1992, ApJ, 396, 161,
  \dodoi{10.1086/171706}

\bibitem[{{Ghisellini} {et~al.}(2010){Ghisellini}, {Tavecchio}, {Foschini},
  {Ghirland a}, {Maraschi}, \& {Celotti}}]{ghisellini_10}
{Ghisellini}, G., {Tavecchio}, F., {Foschini}, L., {et~al.} 2010, MNRAS, 402,
  497, \dodoi{10.1111/j.1365-2966.2009.15898.x}

\bibitem[{{Giannios}(2013)}]{giannios_13}
{Giannios}, D. 2013, MNRAS, 431, 355, \dodoi{10.1093/mnras/stt167}

\bibitem[{{Giannios} \& {Uzdensky}(2019)}]{giannios_uzdensky_19}
{Giannios}, D., \& {Uzdensky}, D.~A. 2019, MNRAS, 484, 1378,
  \dodoi{10.1093/mnras/stz082}

\bibitem[{{Guo} {et~al.}(2014){Guo}, {Li}, {Daughton}, \& {Liu}}]{guo_14}
{Guo}, F., {Li}, H., {Daughton}, W., \& {Liu}, Y.-H. 2014, PhRvL, 113, 155005,
  \dodoi{10.1103/PhysRevLett.113.155005}

\bibitem[{{Kiehlmann} {et~al.}(2016){Kiehlmann}, {Savolainen}, {Jorstad},
  {Sokolovsky}, {Schinzel}, {Marscher}, {Larionov}, {Agudo}, {Akitaya},
  {Ben{\'\i}tez}, {et~al.}}]{kiehlmann_16}
{Kiehlmann}, S., {Savolainen}, T., {Jorstad}, S.~G., {et~al.} 2016, A\&A, 590,
  A10, \dodoi{10.1051/0004-6361/201527725}

\bibitem[{{Larionov} {et~al.}(2013){Larionov}, {Jorstad}, {Marscher},
  {Morozova}, {Blinov}, {Hagen-Thorn}, {Konstantinova}, {Kopatskaya},
  {Larionova}, {Larionova}, {et~al.}}]{larionov_13}
{Larionov}, V.~M., {Jorstad}, S.~G., {Marscher}, A.~P., {et~al.} 2013, ApJ,
  768, 40, \dodoi{10.1088/0004-637X/768/1/40}

\bibitem[{{Loureiro} {et~al.}(2007){Loureiro}, {Schekochihin}, \&
  {Cowley}}]{loureiro_07}
{Loureiro}, N.~F., {Schekochihin}, A.~A., \& {Cowley}, S.~C. 2007, PhPl, 14,
  100703, \dodoi{10.1063/1.2783986}

\bibitem[{{Lyubarsky}(2005)}]{lyubarsky_05}
{Lyubarsky}, Y.~E. 2005, MNRAS, 358, 113,
  \dodoi{10.1111/j.1365-2966.2005.08767.x}

\bibitem[{{Lyutikov} {et~al.}(2018){Lyutikov}, {Komissarov}, {Sironi}, \&
  {Porth}}]{lyutikov_18}
{Lyutikov}, M., {Komissarov}, S., {Sironi}, L., \& {Porth}, O. 2018, JPlPh, 84,
  635840201, \dodoi{10.1017/S0022377818000168}

\bibitem[{{Lyutikov} {et~al.}(2017){Lyutikov}, {Sironi}, {Komissarov}, \&
  {Porth}}]{lyutikov_17}
{Lyutikov}, M., {Sironi}, L., {Komissarov}, S.~S., \& {Porth}, O. 2017, JPlPh,
  83, 635830602, \dodoi{10.1017/S002237781700071X}

\bibitem[{{Lyutikov} \& {Uzdensky}(2003)}]{lyutikov_uzdensky_03}
{Lyutikov}, M., \& {Uzdensky}, D. 2003, ApJ, 589, 893, \dodoi{10.1086/374808}

\bibitem[{{Marscher}(2014)}]{marscher_14}
{Marscher}, A.~P. 2014, ApJ, 780, 87, \dodoi{10.1088/0004-637X/780/1/87}

\bibitem[{{Marscher} {et~al.}(2008){Marscher}, {Jorstad}, {D'Arcangelo},
  {Smith}, {Williams}, {Larionov}, {Oh}, {Olmstead}, {Aller}, {Aller},
  {et~al.}}]{marscher_08}
{Marscher}, A.~P., {Jorstad}, S.~G., {D'Arcangelo}, F.~D., {et~al.} 2008,
  Natur, 452, 966, \dodoi{10.1038/nature06895}

\bibitem[{{Marscher} {et~al.}(2010){Marscher}, {Jorstad}, {Larionov}, {Aller},
  {Aller}, {L{\"a}hteenm{\"a}ki}, {Agudo}, {Smith}, {Gurwell}, {Hagen-Thorn},
  {et~al.}}]{marscher_10}
{Marscher}, A.~P., {Jorstad}, S.~G., {Larionov}, V.~M., {et~al.} 2010, ApJL,
  710, L126, \dodoi{10.1088/2041-8205/710/2/L126}

\bibitem[{{Mehlhaff} {et~al.}(2020){Mehlhaff}, {Werner}, {Uzdensky}, \&
  {Begelman}}]{mehlhaff_20}
{Mehlhaff}, J.~M., {Werner}, G.~R., {Uzdensky}, D.~A., \& {Begelman}, M.~C.
  2020, arXiv e-prints, arXiv:2002.07243

\bibitem[{{Morozova} {et~al.}(2014){Morozova}, {Larionov}, {Troitsky},
  {Jorstad}, {Marscher}, {G{\'o}mez}, {Blinov}, {Efimova}, {Hagen-Thorn},
  {Hagen-Thorn}, {et~al.}}]{morozova_14}
{Morozova}, D.~A., {Larionov}, V.~M., {Troitsky}, I.~S., {et~al.} 2014, AJ,
  148, 42, \dodoi{10.1088/0004-6256/148/3/42}

\bibitem[{{Nalewajko}(2017)}]{nalewajko_17}
{Nalewajko}, K. 2017, Galaxies, 5, 64, \dodoi{10.3390/galaxies5040064}

\bibitem[{{Ortu{\~n}o-Mac{\'\i}as} \& {Nalewajko}(2019)}]{nalewajko_19}
{Ortu{\~n}o-Mac{\'\i}as}, J., \& {Nalewajko}, K. 2019, arXiv e-prints,
  arXiv:1911.06830

\bibitem[{{Petropoulou} {et~al.}(2016){Petropoulou}, {Giannios}, \&
  {Sironi}}]{petropoulou_16}
{Petropoulou}, M., {Giannios}, D., \& {Sironi}, L. 2016, MNRAS, 462, 3325,
  \dodoi{10.1093/mnras/stw1832}

\bibitem[{{Petropoulou} {et~al.}(2019){Petropoulou}, {Sironi}, {Spitkovsky}, \&
  {Giannios}}]{petropoulou_19}
{Petropoulou}, M., {Sironi}, L., {Spitkovsky}, A., \& {Giannios}, D. 2019, ApJ,
  880, 37, \dodoi{10.3847/1538-4357/ab287a}

\bibitem[{{Rowan} {et~al.}(2017){Rowan}, {Sironi}, \& {Narayan}}]{rowan_17}
{Rowan}, M.~E., {Sironi}, L., \& {Narayan}, R. 2017, ApJ, 850, 29,
  \dodoi{10.3847/1538-4357/aa9380}

\bibitem[{{Rowan} {et~al.}(2019){Rowan}, {Sironi}, \& {Narayan}}]{rowan_19}
---. 2019, ApJ, 873, 2, \dodoi{10.3847/1538-4357/ab03d7}

\bibitem[{{Sironi} \& {Beloborodov}(2019)}]{sironi_beloborodov_19}
{Sironi}, L., \& {Beloborodov}, A.~M. 2019, arXiv e-prints, arXiv:1908.08138

\bibitem[{{Sironi} {et~al.}(2016){Sironi}, {Giannios}, \&
  {Petropoulou}}]{sironi_16}
{Sironi}, L., {Giannios}, D., \& {Petropoulou}, M. 2016, MNRAS, 462, 48,
  \dodoi{10.1093/mnras/stw1620}

\bibitem[{{Sironi} {et~al.}(2015){Sironi}, {Petropoulou}, \&
  {Giannios}}]{sironi_15}
{Sironi}, L., {Petropoulou}, M., \& {Giannios}, D. 2015, MNRAS, 450, 183,
  \dodoi{10.1093/mnras/stv641}

\bibitem[{{Sironi} \& {Spitkovsky}(2014)}]{ss_14}
{Sironi}, L., \& {Spitkovsky}, A. 2014, ApJL, 783, L21,
  \dodoi{10.1088/2041-8205/783/1/L21}

\bibitem[{{Sobacchi} \& {Lyubarsky}(2020)}]{sobacchi_20}
{Sobacchi}, E., \& {Lyubarsky}, Y.~E. 2020, MNRAS, 491, 3900,
  \dodoi{10.1093/mnras/stz3313}

\bibitem[{{Spitkovsky}(2005)}]{spitkovsky_05}
{Spitkovsky}, A. 2005, in AIP Conf. Ser., Vol. 801, Simulations of relativistic
  collisionless shocks: shock structure and particle acceleration, ed.
  {T.~Bulik, B.~Rudak, \& G.~Madejski}, 345, \dodoi{10.1063/1.2141897}

\bibitem[{{Uzdensky} {et~al.}(2010){Uzdensky}, {Loureiro}, \&
  {Schekochihin}}]{uzdensky_10}
{Uzdensky}, D.~A., {Loureiro}, N.~F., \& {Schekochihin}, A.~A. 2010, PhRvL,
  105, 235002, \dodoi{10.1103/PhysRevLett.105.235002}

\bibitem[{{Vranic} {et~al.}(2016){Vranic}, {Martins}, {Fonseca}, \&
  {Silva}}]{vranic_16}
{Vranic}, M., {Martins}, J.~L., {Fonseca}, R.~A., \& {Silva}, L.~O. 2016,
  Computer Physics Communications, 204, 141, \dodoi{10.1016/j.cpc.2016.04.002}

\bibitem[{{Werner} \& {Uzdensky}(2017)}]{werner_17}
{Werner}, G.~R., \& {Uzdensky}, D.~A. 2017, ApJL, 843, L27,
  \dodoi{10.3847/2041-8213/aa7892}

\bibitem[{{Werner} {et~al.}(2018){Werner}, {Uzdensky}, {Begelman}, {Cerutti},
  \& {Nalewajko}}]{werner_18}
{Werner}, G.~R., {Uzdensky}, D.~A., {Begelman}, M.~C., {Cerutti}, B., \&
  {Nalewajko}, K. 2018, MNRAS, 473, 4840, \dodoi{10.1093/mnras/stx2530}

\bibitem[{{Werner} {et~al.}(2016){Werner}, {Uzdensky}, {Cerutti}, {Nalewajko},
  \& {Begelman}}]{werner_16}
{Werner}, G.~R., {Uzdensky}, D.~A., {Cerutti}, B., {Nalewajko}, K., \&
  {Begelman}, M.~C. 2016, ApJL, 816, L8, \dodoi{10.3847/2041-8205/816/1/L8}

\bibitem[{{Yuan} {et~al.}(2016){Yuan}, {Nalewajko}, {Zrake}, {East}, \&
  {Blandford}}]{yuan_16}
{Yuan}, Y., {Nalewajko}, K., {Zrake}, J., {East}, W.~E., \& {Blandford}, R.~D.
  2016, ApJ, 828, 92, \dodoi{10.3847/0004-637X/828/2/92}

\bibitem[{{Zenitani} \& {Hoshino}(2001)}]{zenitani_01}
{Zenitani}, S., \& {Hoshino}, M. 2001, ApJL, 562, L63, \dodoi{10.1086/337972}

\bibitem[{{Zhang} {et~al.}(2015){Zhang}, {Chen}, {B{\"o}ttcher}, {Guo}, \&
  {Li}}]{zhang_15}
{Zhang}, H., {Chen}, X., {B{\"o}ttcher}, M., {Guo}, F., \& {Li}, H. 2015, ApJ,
  804, 58, \dodoi{10.1088/0004-637X/804/1/58}

\bibitem[{{Zhang} {et~al.}(2018){Zhang}, {Li}, {Guo}, \& {Giannios}}]{zhang_18}
{Zhang}, H., {Li}, X., {Guo}, F., \& {Giannios}, D. 2018, ApJL, 862, L25,
  \dodoi{10.3847/2041-8213/aad54f}

\end{thebibliography}
\bibliographystyle{aasjournal}


\end{document}